\documentclass[fleqn,10pt]{wlscirep}
\usepackage[utf8]{inputenc}
\usepackage[T1]{fontenc}
\usepackage{tabularx}
\usepackage{booktabs}
\usepackage{fancyhdr}  % Package for headers and footers
\usepackage{draftwatermark}  % For watermark

\SetWatermarkText{Preprint}  % Set the watermark text
\SetWatermarkScale{1.5}  % Scale the size of the watermark
\SetWatermarkColor[gray]{0.9}  % Set the color (light gray)

\title{Exploring the Feasibility of Affordable Sonar Technology: Object Detection in Underwater Environments Using the Ping 360}

\author[1,2,3,*]{Md Junayed Hasan}
\author[1,4]{Somasundar Kannan}
\author[1,4]{Ali Rohan}
\author[5,*]{Mohd Asif Shah}
\affil[1]{National Subsea Centre, Aberdeen AB21 0BH, Scotland, UK}
\affil[2]{School of Social \& Environmental Sustainability, University of Glasgow, Dumfries DG1 4ZL, Scotland, UK}
\affil[3]{Dataxense, Aberdeen AB11 5RG, Scotland, UK}
\affil[4]{School of Computing, Engineering \& Tech., Robert Gordon University, Aberdeen AB10 7AQ, Scotland, UK}
\affil[5]{Department of Economics,Bakhtar University, Kart-e-Char, Kabul, 1001, Afghanistan.}

\affil[*]{junhasan.research@gmail.com}


\begin{abstract}
This study explores the potential of the Ping 360 sonar device, primarily used for navigation, in detecting complex underwater obstacles. The key motivation behind this research is the device's affordability and open-source nature, offering a cost-effective alternative to more expensive imaging sonar systems. The investigation focuses on understanding the behaviour of the Ping 360 in controlled environments and assessing its suitability for object detection, particularly in scenarios where human operators are unavailable for inspecting offshore structures in shallow waters. Through a series of carefully designed experiments, we examined the effects of surface reflections and object shadows in shallow underwater environments. Additionally, we developed a manually annotated sonar image dataset to train a U-Net segmentation model. Our findings indicate that while the Ping 360 sonar demonstrates potential in simpler settings, its performance is limited in more cluttered or reflective environments unless extensive data pre-processing and annotation are applied. To our knowledge, this is the first study to evaluate the Ping 360’s capabilities for complex object detection. By investigating the feasibility of low-cost sonar devices, this research provides valuable insights into their limitations and potential for future AI-based interpretation, marking a unique contribution to the field.
\end{abstract}
\begin{document}

\flushbottom
\maketitle
\thispagestyle{empty}


\section*{Introduction}

In recent years, there has been a growing emphasis on exploring underwater environments. Traditionally, tasks such as inspection, maintenance, and object retrieval were performed by human divers, often at significant physical and mental risk \cite{HP1}. However, advances in technology have led to the development of Remotely Operated Vehicles (ROVs) and Autonomous Underwater Vehicles (AUVs), which have transformed ocean exploration by automating complex tasks and reducing human risks \cite{HP2}. Despite these advancements, underwater navigation and object detection remain challenging due to the limitations of optical sensors. Water turbidity, light attenuation, and scattering severely degrade visual data quality, making traditional cameras unreliable in sub-sea operations \cite{HP3}. While short-range improvements in visual data quality have been achieved, alternative sensing technologies are needed to overcome these limitations \cite{HP6}.

Sonar technology has emerged as a viable solution for underwater detection and navigation, effectively penetrating murky waters and low-light conditions through the use of sound waves \cite{HP8}\cite{HP9}. There are two primary types of sonar systems: single-beam and multi-beam. Multi-beam sonar systems, which generate detailed 3D maps and provide extensive spatial coverage, offer superior resolution and accuracy. However, their high cost limits accessibility for many applications \cite{HP11}. The Ping 360 sonar, developed by Blue Robotics, presents an affordable alternative as a single-beam system designed primarily for navigation. Capable of performing 360° scans, it offers potential for underwater exploration and obstacle avoidance \cite{HP12}. However, its ability to perform more advanced tasks, such as complex object detection, remains under explored. Accurate object detection is critical for autonomous underwater vehicles operating without human intervention, where the reliable identification of underwater structures is essential. This study aims to investigate the feasibility of using the Ping 360 sonar for underwater object detection in real-world scenarios.

Existing research has explored sonar-based object detection, particularly through imaging sonar techniques. However, these approaches often fail to account for the fundamental differences between sonar and optical imagery. Single-beam sonar systems like the Ping 360 generate acoustic representations of underwater environments, which are sometimes converted into image-like outputs. While this conversion can improve detection accuracy, it oversimplifies the inherent complexities of sonar data, such as non-homogeneous resolution, speckle noise, acoustic shadowing, and reverberation, as noted by Karimanzira et al. \cite{HP13}. While the Ping 360 is an affordable option—about one-fifth the cost of multi-beam systems—it faces challenges such as noise interference and shadowing, which complicate object detection \cite{HP10}. Despite these limitations, single-beam systems are practical for smaller-scale or budget-conscious operations. Kim et al. suggest that advanced image processing techniques can help mitigate these limitations, though they may still struggle in noisy or low-resolution environments \cite{HP14}. In more complex underwater environments, such as turbid or structurally intricate coastal habitats, the resolution of the Ping 360 cannot match that of high-frequency multi-beam sonars, which offer real-time, camera-like visuals. However, McKay et al. showed that fine-tuning machine learning models for sonar data can improve detection accuracy, even with simpler systems like the Ping 360 \cite{HP13}. Still, these models often rely on high-quality sonar data, which is not always realistic for single-beam systems. Our study addresses these challenges by evaluating the Ping 360 sonar’s potential for underwater object detection in real-world settings, moving beyond its traditional navigation role. 

Many existing studies focus on imaging sonar or synthetic data generation, such as Jiang et al.'s CycleGAN approach, which converts sonar data into image-like formats \cite{HP15}. While these methods produce strong results in controlled environments, they often overlook the complexities of sonar data, such as variable resolution, acoustic noise, and shadowing. Forcing sonar data into image-based frameworks risks misinterpreting its unique characteristics, which are crucial for real-world detection accuracy. Recent studies frequently apply image processing and AI models to sonar data \cite{HP17}\cite{HP18}\cite{HP19}\cite{HP20}. While these models may perform well in controlled environments, they often struggle in real-world conditions where sonar’s acoustic properties must be accounted for. Moreover, manual annotation processes are often misrepresented, leading to errors in sonar data interpretation and artificially inflated model performance metrics, as noted by Zhao et al. \cite{HP15}. Factors such as object material, size, and positioning, which significantly impact detection reliability, are rarely considered in detail.

Our study focuses on re-purposing the Ping 360; the navigation-sonar; for more complex underwater object detection tasks. Rather than developing AI solutions, we evaluate the device’s feasibility and limitations in environments where traditional image-based methods fail, such as in low-visibility or turbid waters. We emphasize the importance of treating, and interpreting sonar data based on its acoustic properties, rather than forcing it into an image-processing framework. By examining challenges such as surface reflections, noise, and acoustic shadows, we provide a realistic evaluation of the Ping 360’s capabilities for complex detection tasks. This research, being the first of its kind, offers a unique contribution to the field.

\subsection*{Research Questions}
The primary research questions (RQ) addressed in this study are as follows:
\begin{itemize}
\item	\textbf{RQ1:} What are the main challenges in interpreting navigation sonar data in cluttered or reflective environments?
\item	\textbf{RQ2: }How effective is manual annotation of sonar data in improving object detection when used with advanced segmentation AI models such as UNet?
\item	\textbf{RQ3:} Can low-cost sonar devices like Ping 360 typically used for navigation, be effectively repurposed for complex underwater object detection?
\end{itemize}

\subsection*{Contributions}
In line with finding out the answers to those questions, this research makes several key contributions:
\begin{itemize}
\item	For the first time, we present a detailed investigation into the feasibility of using the Ping 360 sonar device for object detection in underwater environments. 
\item	We are releasing the raw collected data from the Ping 360 sonar to enable future researchers to explore its potential usability, especially for those who may not have access to the device due to its cost, despite it being relatively affordable.
\item	We provide insights into the challenges of sonar data interpretation, including surface reflections, object shadows, and the importance of careful data annotation for machine learning.
\item	We developed a segmentation dataset from manually annotated sonar images and demonstrated that the Ping 360 sonar's object detection performance is highly dependent on data pre-processing and careful interpretation, particularly when used with machine learning models like the U-Net segmentation algorithm.
\end{itemize}

\section*{Experimental set-up details}

In this section, we present the details of the experimental setup used in this study. The primary focus is to explain the scanning mechanism of the Ping 360 and how we configured the environment for data collection.

\begin{figure}[ht]
\centering
\includegraphics[width=0.5\linewidth]{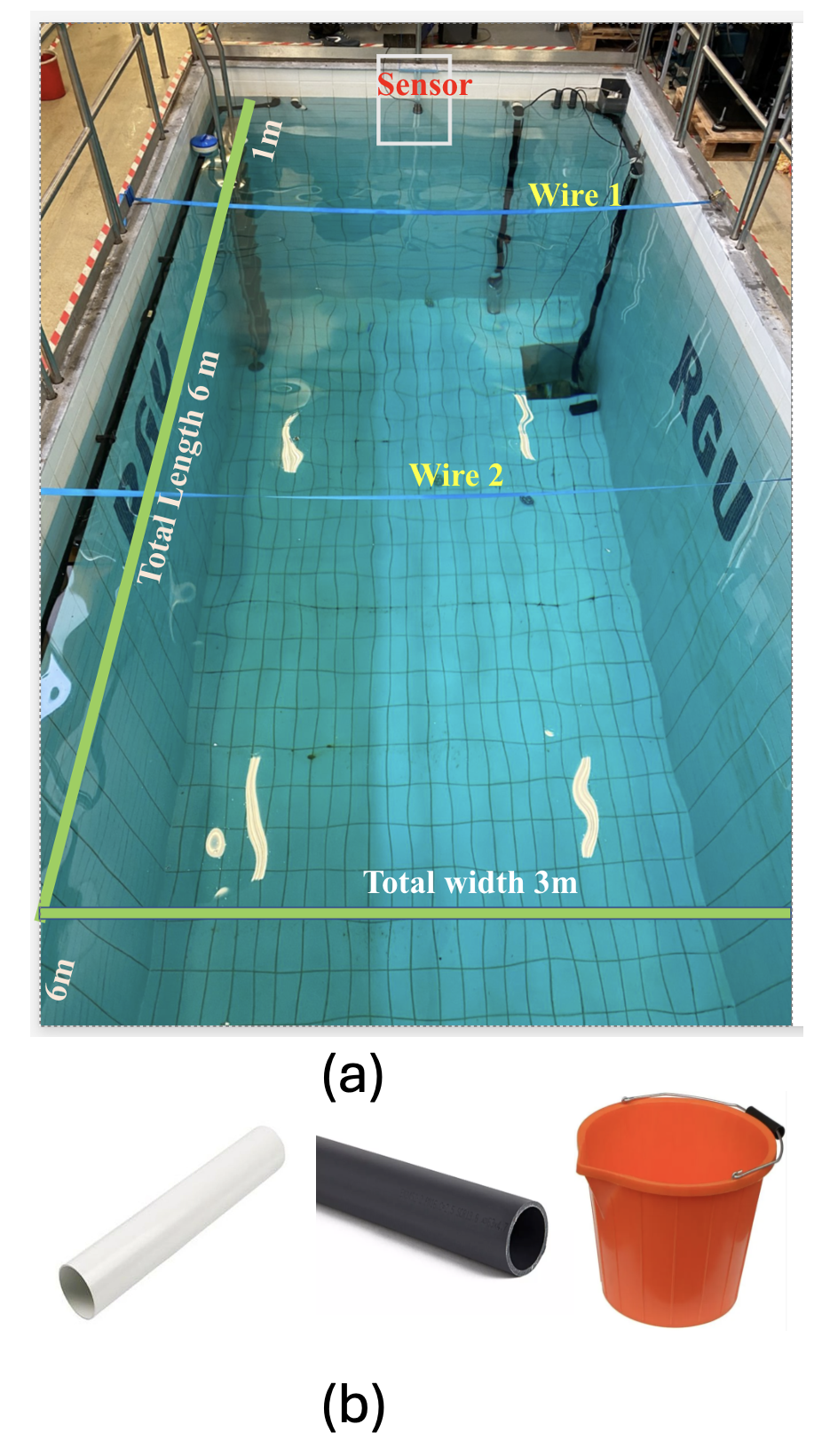}
\caption{(a) Experimental set up of the small pool, (b) used objects for sonar scan}
\label{fig:1}
\end{figure}

The experiment was conducted in a small pool with a total width of 3 meters, a length of 6 meters, and depth of 5 meters. The sonar was positioned 1.5 meters from the side of the pool, placing it exactly at the middle of the pool's width. The sonar sensor was oriented forward along the pool, with its transducer head aligned at 0° relative to the x-axis. The sonar was set to perform a forward-facing scan, covering a 180° sector, scanning from -90° to +90° relative to the x-axis. This sonar has a horizontal beam width of 1° to 2°, and a vertical beam height of 20° to 30° \cite{HP12}. These narrow horizontal and wide vertical fields of view allowed us to focus on precise horizontal slices of the underwater environment, which is essential for detecting objects at various depths.

In the pool, two wires were positioned at different distances along the length to hang various objects, simulating the offshore floating structures. The wires served as attachment points for objects, and their setup made it easier to change and remove the objects as needed throughout the experiment. Furthermore, since the exact positions of the wires were known, we had ground truth data for the range of the objects relative to the sonar (straight distance - x -axis). This allowed us to verify the accuracy of the sonar data in measuring range or  even detecting objects at known distances from the side of the pool end where the sonar was mounted. The wires were placed as follows; wire 1 was placed at 2 meter from the sensor, and wire 2 was positioned 4 meter from the sensor (Fig.~\ref{fig:1}).  This setup allowed us to track object positions (e.g., pvc grey and white pipe, orange bucket) and measure the sonar’s accuracy in detecting and localizing the objects as they were moved between different points as explained before.

\begin{figure}
    \centering
    \includegraphics[width=1\linewidth]{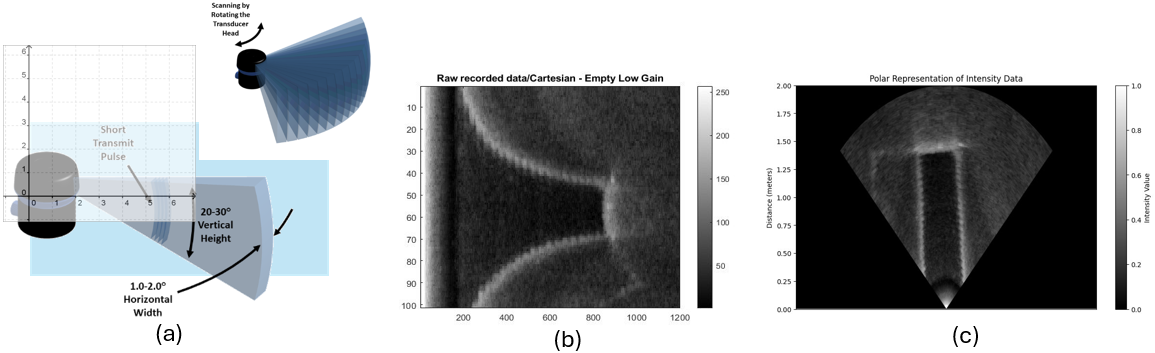}
    \caption{(a) Illustration of the sonar sensor positioned at 0° horizontal, with its transducer head aligned along the x-axis. The expected vertical and horizontal coverage of the transmitted pulse is also shown\cite{HP12}, along with the scan area (represented by the blue rectangular box) in a hypothetical tank; (b)Raw sonar output data from a 102° scan of an empty tank with a boundary at approximately 1.50 meters. The left axis shows the intensity values (I) ranging from 0 to 255 in a polar coordinate system;  (c) Converted polar representation of the raw data in (b), where the tank boundary is visible at 1.50 meters. This form is typically displayed by the sonar interface software and is commonly used in AI algorithm-driven research for sonar data analysis. All figures are shown for demonstration purposes. }
    \label{fig:2}
\end{figure}

The sonar scanned the area in small increments, with each angle's data recorded as intensity values \(I\), ranging from 0 to 255. These intensity values represent the strength of the sound wave reflections from objects in the pool. In this experiment, we set the scanning range to 7 meters to ensure full coverage of the pool boundary, which extends beyond the maximum pool length of 6 meters. This extra meter provides a buffer zone for detecting objects or boundaries near the pool's edge. Considering the sample length of 1200, each data point corresponded to approximately 0.00583 meters, balancing the need for accuracy while working within the device's sampling constraints. During data collection, the sonar's scanning depth was set at 0.15 meters below the surface of the water, and the scanning was performed at a water temperature of 10°C. The water salinity was set to 0 psu to simulate freshwater conditions. The sonar was operated at a medium gain setting (G1). It is important to note that, during operation, the Ping 360 sonar generates significant noise near the sonar head due to the rotational motion of the transducer. This noise creates approximately 0.75 meters of disturbance surrounding the sonar \cite{HP12}. In Figure 2, we demonstrate the sonar's position (a), the raw data output (b), and the converted output typically displayed by the sonar interface software that comes with the sensor purchase, which is also commonly used in recent AI algorithm-driven research.

\section*{Relevant Technical Details}

In this section, we will discuss the details about the sonar propagation, the sonar data pre-processing methodology, and proposed architecture of U-Net algorithm.

\subsection*{Underwater Sound Propagation}

In underwater environments, the propagation of sound behaves quite differently from its behaviour in air. One of the key differences is the variability of the speed of sound, \(c\) (measured in m/s), which is not constant underwater \cite{HP26}\cite{DP3}. The speed of sound is influenced by several factors, including water temperature \(T\) (in °C), depth \(D\) (in meters), and salinity \(S\) (in practical salinity units, psu). This relationship is expressed by the following equation:

\begin{equation}
\label{eq:sound}
c = 1410 + 4.21 \times T - 0.037 \times T^2 + 1.1 \times S + 0.018 \times D
\end{equation}

In this study, as we are conducting experiments in freshwater environments, the salinity \(S\) is consistently set to 0 psu. Understanding and calculating the speed of sound underwater is essential because it directly impacts the performance and range of sonar systems. The system's range is particularly dependent on the speed of sound, as well as additional parameters such as the sample period \(S_p\) (in seconds), sample distance \(S_d\) (in meters), and the total number of samples \(S_n\). For this sonar device, the maximum scan range is not directly set by the user but is instead determined by these environmental and system parameters. The following equations explain the relationships involved:

\begin{equation}
\label{eq:Sd}
S_d = \frac{c \times S_p}{2}
\end{equation}

\begin{equation}
\label{eq:maxrange}
\text{Max Range} = S_d \times S_n
\end{equation}

Equation~(\ref{eq:Sd}) calculates the sample distance \(S_d\) based on the speed of sound \(c\) and the sample period \(S_p\). This distance is then multiplied by the number of samples \(S_n\) to determine the sonar’s maximum range, as shown in equation~(\ref{eq:maxrange}). These relationships are critical for configuring the Ping 360 sonar to perform optimally under various environmental conditions. By adjusting the speed of sound based on the water temperature and depth, and carefully configuring parameters such as the sample period and number of samples, we can achieve better resolution of data.

\subsection*{Sonar signal processing}

Initial data filtering excluded unreliable points within 0.75 meters of the sonar head, as recommended by Bluerobotics \cite{HP12}, and beyond 6 meters, which corresponded to the full length of the pool and the maximum effective scanning range for the experimental setup. This filtering ensured that only valid data points within the useful scanning range were retained for analysis. Following this, a statistical threshold approach was applied to further refine the region of interest (ROI) \cite{DP2}. Intensity values \(I\) for each scan angle were retained only if they met the condition \(I \geq 2 \times \mu + \sigma\), where \(\mu\) is the mean intensity and \(\sigma\) is the standard deviation. This approach ensured the removal of noise while preserving significant reflections from objects in the pool environment. The mean intensity \(\mu\) and standard deviation \(\sigma\) for each scan angle were calculated using the following equations:

\begin{equation}
\mu = \frac{1}{N'} \sum_{i=1}^{N'} I_i
\label{eq:4}
\end{equation}

\begin{equation}
\sigma = \sqrt{\frac{1}{N'-1} \sum_{i=1}^{N'} (I_i - \mu)^2}
\label{eq:5}
\end{equation}

where \(N'\) is the number of samples within the defined ROI.

Once the noise was filtered out, the denoised data was converted from Cartesian to Polar coordinates for easier analysis. This conversion allowed for more accurate interpretation of the sonar data in terms of the distances and angles relative to the sonar head. The transformation from Cartesian coordinates \((x, y)\) to polar coordinates \((r, \theta)\) is described by the following equations:

\begin{equation}
r = \sqrt{x^2 + y^2}
\label{eq:6}
\end{equation}

\begin{equation}
\theta = \arctan\left(\frac{y}{x}\right)
\label{eq:7}
\end{equation}

\subsection*{Segmentation Algorithm - U-Net}

U-Net is a convolution neural network architecture designed for image segmentation, particularly useful in scenarios where precise pixel-wise classification is needed. The architecture is composed of a contracting path (encoder) followed by an expansive path (decoder). The contracting path captures context by progressively downsampling the input, while the expansive path restores spatial resolution by upsampling. Skip connections are used between corresponding layers in the encoder and decoder to combine feature maps, ensuring that both context and localization information are preserved \cite{HP27}. The key goal of U-Net in this study is to segment sonar images by distinguishing object regions from background noise. In this study, after analyzing the data from the sonar, which was collected in Cartesian coordinates, the data is manually inspected and transformed into polar co-ordinates and converted into a grey-scale image format for training the U-Net. With this types of input, the model learns to separate objects from the background based on intensity variations detected by the sonar. Table ~\ref{tab:1} highlights the details of the proposed U-Net architecture in this study.

\begin{table}[ht]
\centering
\caption{U-Net structure for grayscale sonar image segmentation}
\label{tab:1}
\begin{tabular}{lccccc}
\toprule
\textbf{Layer Type}         & \textbf{Input Size} & \textbf{Output Size} & \textbf{Filter Size} & \textbf{Activation} \\ 
\midrule
Input                       & (1, H, W)           & (1, H, W)            & -                    & -                   \\ 
Conv2D                      & (1, H, W)           & (64, H, W)           & 3x3                  & ReLU                \\ 
Conv2D                      & (64, H, W)          & (64, H, W)           & 3x3                  & ReLU                \\ 
MaxPooling                  & (64, H, W)          & (64, H/2, W/2)       & 2x2                  & -                   \\ 
Conv2D                      & (64, H/2, W/2)      & (128, H/2, W/2)      & 3x3                  & ReLU                \\ 
Conv2D                      & (128, H/2, W/2)     & (128, H/2, W/2)      & 3x3                  & ReLU                \\ 
MaxPooling                  & (128, H/2, W/2)     & (128, H/4, W/4)      & 2x2                  & -                   \\ 
Conv2D                      & (128, H/4, W/4)     & (256, H/4, W/4)      & 3x3                  & ReLU                \\ 
Conv2D                      & (256, H/4, W/4)     & (256, H/4, W/4)      & 3x3                  & ReLU                \\ 
MaxPooling                  & (256, H/4, W/4)     & (256, H/8, W/8)      & 2x2                  & -                   \\ 
Conv2D                      & (256, H/8, W/8)     & (512, H/8, W/8)      & 3x3                  & ReLU                \\ 
UpSampling                  & (512, H/8, W/8)     & (512, H/4, W/4)      & -                    & -                   \\ 
Concat (Skip Connection)     & (512+256, H/4, W/4) & (768, H/4, W/4)      & -                    & -                   \\ 
Conv2D                      & (768, H/4, W/4)     & (256, H/4, W/4)      & 3x3                  & ReLU                \\ 
UpSampling                  & (256, H/4, W/4)     & (256, H/2, W/2)      & -                    & -                   \\ 
Concat (Skip Connection)     & (256+128, H/2, W/2) & (384, H/2, W/2)      & -                    & -                   \\ 
Conv2D                      & (384, H/2, W/2)     & (128, H/2, W/2)      & 3x3                  & ReLU                \\ 
UpSampling                  & (128, H/2, W/2)     & (128, H, W)          & -                    & -                   \\ 
Concat (Skip Connection)     & (128+64, H, W)      & (192, H, W)          & -                    & -                   \\ 
Conv2D                      & (192, H, W)         & (64, H, W)           & 3x3                  & ReLU                \\ 
Conv2D (Final)              & (64, H, W)          & (1, H, W)            & 1x1                  & Sigmoid             \\ 
Output                      & (1, H, W)           & (1, H, W)            & -                    & -                   \\ 
\bottomrule
\end{tabular}
\end{table}

For each input sonar image, let the input image be represented as \( X \in \mathbb{R}^{H \times W} \), where \(H\) and \(W\) are the height and width of the image, respectively. The U-Net performs segmentation by minimizing a loss function that measures the discrepancy between the predicted output \( Y_{\text{pred}} \) and the ground truth labels \( Y_{\text{true}} \). A common loss function used is the binary cross-entropy loss:

\begin{equation}
\mathcal{L}_{\text{BCE}} = - \frac{1}{N} \sum_{i=1}^{N} \left( Y_{\text{true}, i} \log (Y_{\text{pred}, i}) + (1 - Y_{\text{true}, i}) \log (1 - Y_{\text{pred}, i}) \right)
\label{eq:8}
\end{equation}

where \(N\) is the total number of pixels in the image, \(Y_{\text{true}, i}\) represents the ground truth for pixel \(i\), and \(Y_{\text{pred}, i}\) is the predicted probability for pixel \(i\) being part of the object. Additionally, the Dice coefficient, which measures the overlap between the predicted segmentation and the ground truth, is often used to evaluate segmentation performance. The Dice loss can be defined as:

\begin{equation}
\mathcal{L}_{\text{Dice}} = 1 - \frac{2 \sum_{i=1}^{N} Y_{\text{true}, i} Y_{\text{pred}, i}}{\sum_{i=1}^{N} Y_{\text{true}, i} + \sum_{i=1}^{N} Y_{\text{pred}, i}}
\label{eq:9}
\end{equation}

\section*{Dataset Details}

\begin{table}[ht]
\centering
\caption{Experimental details for collecting sonar data}
\label{tab:2}
\begin{tabular}{|c|p{3cm}|p{5cm}|p{6 cm}|}
\toprule
\textbf{Exp.} & \textbf{Object Position(m)} & \textbf{Object Types}  & \textbf{Remarks} \\
\midrule
1  & None             & None                          & \\ \hline 
2  & 2                & Grey PVC pipe                 & \\ \hline 
3 - 5 & 2, 4            & Grey and white PVC pipes      & The placement of grey and white PVC pipes at wire 1 and wire 2 is unique in each experiment. \\ \hline 
6  & 2, 2             & Grey and white PVC pipes      & \\ \hline 
7  & 2, 2, 4          & Grey and white PVC pipes, orange bucket & The placement of grey and white PVC pipes at wire 1 is different. \\ \hline 
8  & 2, 4             & White PVC pipe, orange bucket & \\ \hline 
9  & 4                & Orange bucket                 & \\ \hline 
10& 2, 4          & 2 orange buckets              & \\ \hline
\end{tabular}
\end{table}

The dataset consists of sonar data in its original format collected from 10 experimental scenarios using the Ping 360 sonar device. The purpose of these experiments was to observe the Ping 360's capability to detect objects placed at varying distances from the sensor in a controlled pool environment. The data were captured using a consistent sonar configuration across all experiments, with the gain set to medium (G1) and 1200 samples collected per scan. In these experiments, objects were attached to two wires stretched across the width of the pool (which measures 3 meters). Wire 1 was positioned 2 meters from the sensor and Wire 2 was positioned 4 meters away. The objects were hooked at various positions along the length of the wires, allowing for variable placements within the pool’s 3-meter width. This setup simulates different object positions horizontally and vertically, offering a realistic representation of object distribution in an underwater environment. The objects used in the experiments included two PVC pipes (one grey and one white) and two orange buckets. The first experiment was conducted with no objects present, serving as a baseline. Subsequent experiments introduced objects in different configurations and numbers, allowing for analysis of how object type, position, and density affect sonar detection. In the following Figures.~\ref{fig:3}, and \ref{fig:4} the experimental setup and the collected occupancy polar map are shown for qualitative analysis in the subsequent section.
\begin{figure}
    \centering
    \includegraphics[width=1\linewidth]{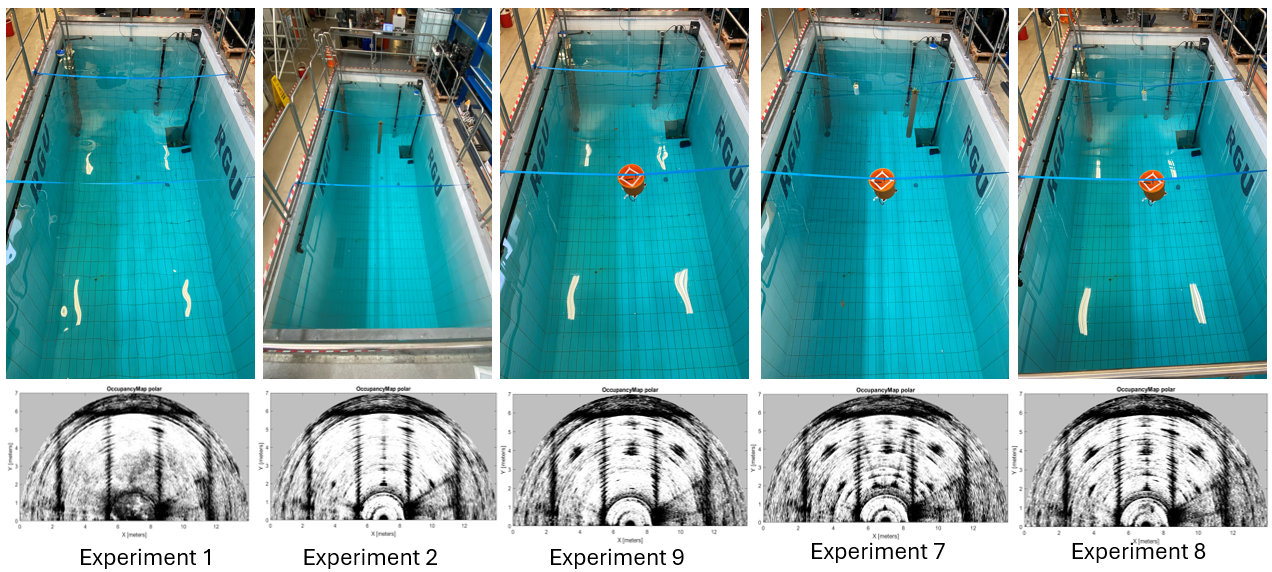}
    \caption{Experimental setup pictures and corresponding Ping 360 software - ping viewer processed polar occupancy map of the collected data}
    \label{fig:3}
        \centering
    \includegraphics[width=1\linewidth]{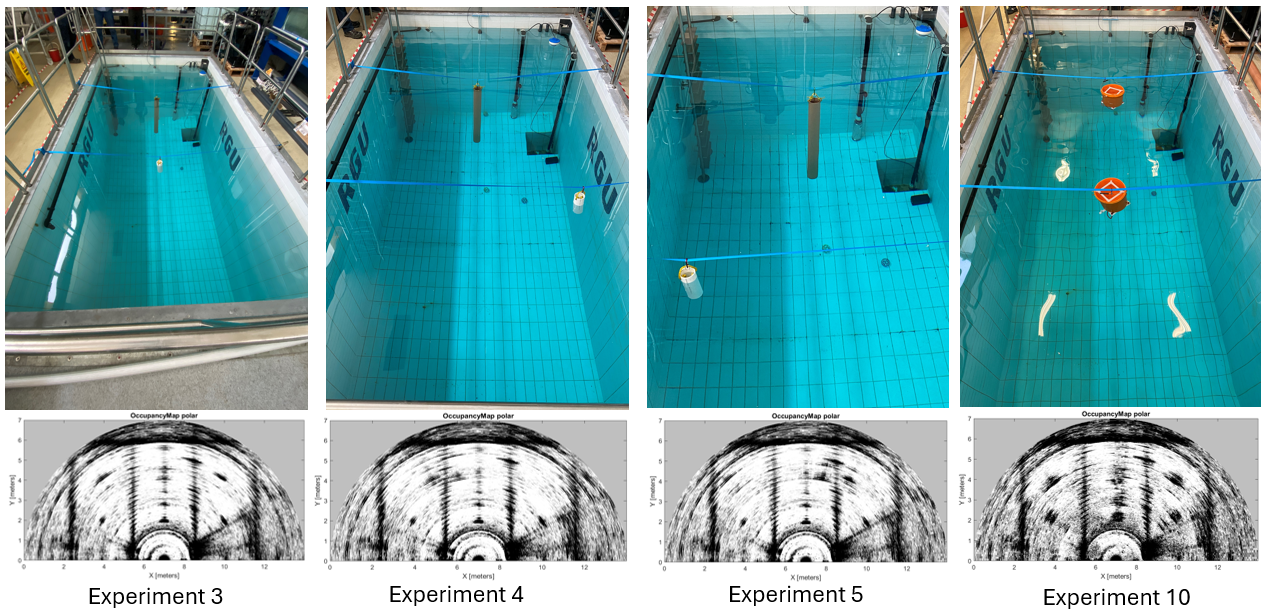}
    \caption{Experimental setup pictures same as above but tailored for understanding acoustic shadow}
    \label{fig:4}
\end{figure}

\section*{Experimental Analysis}
\subsection*{Understanding Polar Occupancy Map}

\begin{figure}
    \centering
    \includegraphics[width=1\linewidth]{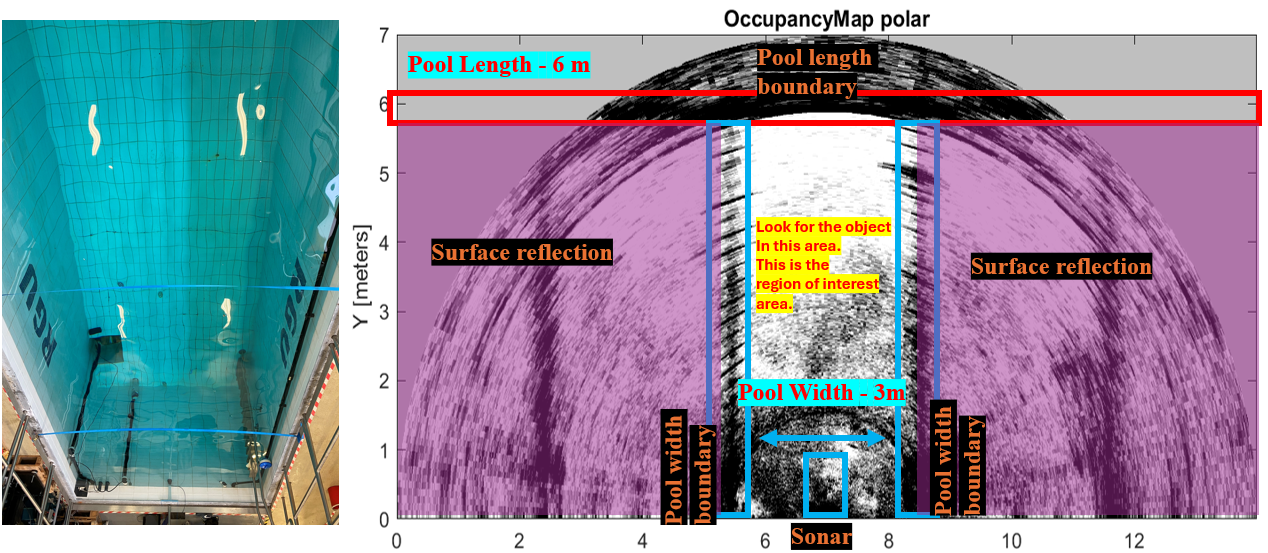}
    \caption{Experimental set up of experiment 1, and enhanced polar occupancy map}
    \label{fig:5}
    \includegraphics[width=0.75\linewidth]{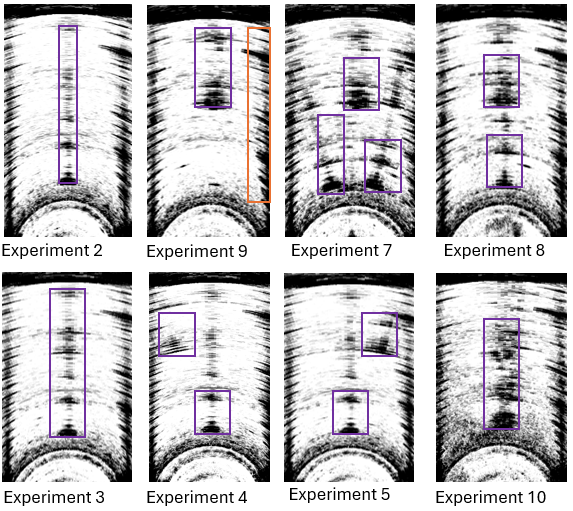}
    \caption{Enhanced polar occupancy map considering ROI for all the experiments}
    \label{fig:6}
\end{figure}

Before analysing the data presented in Figures~\ref{fig:3} and \ref{fig:4}, it is essential to explain how to interpret the occupancy map and define the region of interest (ROI) based on both subjective understanding and experimental pre-knowledge. As illustrated in Figure~\ref{fig:5}, the experimental details from Experiment 1 have been enhanced, and the photo of the setup is shown upside down to match the viewing perspective of the occupancy polar map.

In the occupancy polar map, the pool boundaries, including both height and width, are clearly displayed. The blue rectangular shape highlights the boundary bars visible in the occupancy map. The distance between the left and right pool width boundaries is 3 meters, which is also confirmed by the Ping Viewer software-generated occupancy map. The pool length boundary is at 6 meters, and this is also marked by a strong boundary, enclosed within a red box. Beyond this 6-meter mark, surface reflections begin to appear. A similar phenomenon can be observed after marking the pool width boundaries, where the light purple zone denotes the surface reflection area. In this setup, the sonar was positioned 1.5 meters from the side of the pool, placed precisely in the centre of the pool's width. The sonar sensor was oriented forward along the pool, with its transducer head aligned at 0° relative to the x-axis, and it was not submerged deeply in the water. Additionally, the pool is constructed with ceramic tiles, a reflective material that contributes to the surface reflections. When inspecting objects detected by sonar and their reported ranges, it is crucial to focus on the ROI, ensuring the most accurate readings. According to the data-sheet of the Ping 360 sonar, the minimum working range is 0.75 meters \cite{HP12}. Therefore, within the ROI, there is a significant amount of noise (indicated by darker colours) from 0.75 meters to 1 meter. Moreover, some noise extends beyond this range, due to surface reflections. The pool is not entirely empty; it contains setup pipes and other objects that generate passive noise, further affecting the sonar readings. Therefore, when analysing the data, special attention must be given to these factors to accurately interpret the results.

\subsection*{Qualitative Inspection of the Polar Occupancy Map}

Let's now analyse the experimental data presented in Figures~\ref{fig:3} and \ref{fig:4}. For a more refined analysis, the data from each experiment has been thresholded based on the determined ROI, focusing on the range where meaningful data exists. Any areas outside this range, particularly those beyond the ROI, have been discarded to avoid unnecessary noise and interference. Furthermore, the first 1 meter from the starting position of the sonar scan has been removed due to significant noise in that zone, as supported by the Ping 360 sonar data-sheet. The qualitative analysis here is based on Figures~\ref{fig:3}, \ref{fig:4}, \ref{fig:6}, and Table~\ref{tab:2}.

\subsubsection*{Experiment 2: Pipe Detection and Acoustic Shadow}
As seen in Figures~\ref{fig:3}, \ref{fig:6}, and Table~\ref{tab:2}, the starting point of the pipe in Experiment 2 is detected at the 2-meter mark. Both visual observation and ground truth confirm this detection. However, following the detection, a long black line appears in the sonar image, highlighted within the purple rectangle in Figure~\ref{fig:6} (Experiment 2). This is a clear instance of acoustic shadow. Here, the detected pipe blocks the sonar's sound waves, preventing them from reaching the area behind it, thus creating this shadowed region devoid of sonar returns.

\subsubsection*{Experiment 9: Bucket Detection and Acoustic Shadow}
In Experiment 9, we are dealing with an orange bucket whose width is significantly larger than that of the grey PVC pipe used in Experiment 2. As a result, the bucket is detected more strongly, and a pronounced acoustic shadow forms behind it, as seen in the purple rectangle of Figure~\ref{fig:6}. However, due to the bucket's smaller longitudinal dimension compared to the pipe, the acoustic shadow is shorter than what we observed with the PVC pipe. The length of the object, in this case, plays a direct role in determining the size of the acoustic shadow, as a longer object blocks more sound waves, casting a larger shadow. Interestingly, surface reflections also play a part in creating acoustic shadows, as highlighted in the orange annotation in Experiment 9 of Figure~\ref{fig:6}. This shows that the pool's reflective surface can sometimes contribute to the shadowing effect, in addition to the objects present in the sonar’s path.

\subsubsection*{Experiment 8: Combined Object Setup}
In Experiment 8, a bucket and a small white PVC pipe were placed in the same straight line, with the pipe in wire 1 and the bucket in wire 2. Based on the sonar setup, we would expect the 2-meter object (the PVC pipe) to be detected clearly, while the object at 4 meters (the bucket) should be partially obscured by the acoustic shadow of the first pipe, yet still somewhat detectable because the bucket is wider. This hypothesis was confirmed during the experiment, as the bucket, despite being behind the first object, was still detectable. Some reflections from the second object may have been observed, but they did not pose significant difficulties in distinguishing or detecting the objects. The larger width of the bucket helped mitigate the shadow effect to some extent, which is in line with our understanding of how object size influences the severity of the shadow.

\subsubsection*{Experiment 7: No Acoustic Shadow Interference}
In Experiment 7, grey and white PVC pipes were placed on wire 1 with a gap between them, and an orange bucket was placed on wire 2, aligned with the gap. Since none of the objects directly blocked one another, each was clearly detected without being hidden under an acoustic shadow. The absence of overlapping objects in the sonar's line of sight allowed for a clean detection of all items, without interference from the shadowing effects.

\subsubsection*{Experiment 3: Complete Acoustic Shadow}
In contrast to Experiment 8, Experiment 3 involved two PVC pipes with the same width placed in different wires. Here, the acoustic shadow cast by the first pipe completely obscured the second pipe, which was aligned behind it on wire 2. Since both pipes had the same width, the acoustic shadow of the first object fully covered the second, resulting in its non-detection. This experiment underscores how object alignment and size play critical roles in the formation of acoustic shadows and how they affect sonar detection capabilities.

\subsubsection*{Experiment 10: Strong Detection of Bucket and Shadow}
Experiment 10 exhibited similar behaviour to Experiment 9, where a larger object, in this case, a bucket, was placed in front of PVC pipes. The detection of the bucket was strong, as its width exceeded that of the pipes, and it cast a well-defined acoustic shadow behind it. However, because of the bucket's dimensions, the shadow was not as long as in cases where longer objects, like pipes, were detected. This once again demonstrates the correlation between an object's size and the extent of the acoustic shadow it produces.

\subsubsection*{Experiments 4 and 5: No Acoustic Shadow or Reflection Issues}
In Experiments 4 and 5, no issues were encountered with acoustic shadows or surface reflections, allowing for clear detection of objects at both distances. The sonar was able to capture the objects without interference, as no objects were positioned in a way that would block sound waves and create shadowed areas. The absence of significant reflections from the pool's surface or nearby objects further contributed to the clean detection in these experiments.

\subsubsection*{Observation Summary}
Through these experiments, the impact of acoustic shadows in sonar imagery is clearly demonstrated. The presence of an object directly blocks sonar waves, preventing them from reaching areas behind the object and creating shadowed regions where no returns are detected. The size, shape, and alignment of objects are critical factors in determining the extent and intensity of these shadows. Additionally, surface reflections from the pool can contribute to or amplify acoustic shadow effects, particularly in complex setups involving multiple objects. This understanding is crucial for accurate sonar data interpretation and for mitigating the effects of acoustic shadows in practical applications.

\subsection*{Challenges in Sonar Data Annotation: Addressing Ambiguities and Operational Factors}

When defining objects in sonar imagery, as shown in the occupancy maps in Figures~\ref{fig:3}, ~\ref{fig:4}, and particularly in Figure~\ref{fig:6}, where the ROI-based polar occupancy map highlights objects with bounding boxes, several challenges emerge. Take, for instance, Experiment 8 in Figure~\ref{fig:6}; can we confidently state that the bounding boxes represent a PVC pipe and a bucket? Not necessarily. While we can infer the presence of potential obstacles, accurately distinguishing between different object types is difficult without additional context. This highlights one of the key challenges in sonar-based object detection: interpreting ambiguous or unclear sonar data.

\subsubsection*{Challenges in Object Annotation}
In scenarios like this, when annotating data for models such as U-Net-based object detection, it becomes essential to incorporate pre-knowledge or human supervision. The visible sonar image alone may not provide enough information to confidently label an object. Thus, experts, such as ROV/AUV pilots who have years of experience navigating underwater environments, play a critical role. They bring domain knowledge that helps guide the annotation process, suggesting what specific features might represent. This expert input, combined with the sonar data, enables more accurate bounding box placements and labeling, reducing the margin for error that typically arises when interpreting sonar imagery. For instance, in the provided image, the two bounding boxes indicate areas of interest. However, without external knowledge, it is difficult to discern whether the objects are PVC pipes, buckets, or other obstructions. Here, a domain expert would likely know that specific patterns or shapes in the data correspond to certain object types, which would be invaluable during the annotation process.

\subsubsection*{Impact of Sensor Position and Orientation}
Another critical factor is the position and orientation of the sonar sensor. In this case, the sonar sensor is positioned at 0° horizontal, with its transducer head aligned along the x-axis. While this setup provides one view of the underwater environment, it limits what can be seen. If the sensor were angled differently—such as being tilted downward like a torchlight beam—it might reveal more distinct shapes or features of objects. However, the visibility of such shapes would still depend heavily on the operating conditions, such as water clarity, object material, and the acoustic properties of the environment. For instance, a downward-facing sensor might reveal more detail about the height or vertical structures of objects in the water, which would be harder to detect in a purely horizontal configuration. The operating angle and position can drastically affect the quality of the sonar image and, consequently, the accuracy of object detection and annotation.

\subsubsection*{Gaps in Existing Research}
In the introduction, we referenced various studies that apply image processing techniques and AI models to sonar data for object detection \cite{HP17}\cite{HP18}\cite{HP19}\cite{HP20}\cite{HP21}\cite{HP22}\cite{HP23}\cite{HP24}\cite{HP25}. While these models often report high accuracy in controlled test datasets, they tend to overlook real-world complexities such as the acoustic shadow effects, cluttered environments, or reflective surfaces. These studies primarily focus on pixel-level precision in sonar imagery, yet rarely discuss the deeper challenges of sonar data interpretation and annotation. The annotation process in these studies is often assumed, but critical details such as whether subject matter experts contributed to the process, the mounting angle of the sonar, or how the ROV/AUV’s movement and positioning altered the sonar angle and subsequently affected the collected data are frequently left unexplained. Without accounting for these factors, the reported high accuracy of these models can be misleading when applied to more complex or real-world environments, where ambiguity in sonar data is common.

Therefore, in the next section, we will proceed with annotating the data based on the observations we have made. Specifically, we will create segmentation masks for the detected objects, categorizing each mask as an object class. Using these annotations, we will evaluate the performance of the U-Net model. According to our hypothesis, by applying the data processing techniques discussed earlier—such as noise reduction—we anticipate an improvement in accuracy. Additionally, increasing the training duration or initializing the model with well-tuned weights could further enhance the results. While more complex models, as discussed in the literature, might yield even higher accuracy.

\section*{Result Analysis of Segmentation Model}

To evaluate the performance of our U-Net-based object detection model, we collected approximately 700 samples across various experimental scenarios. Each sample was manually annotated, creating segmentation masks by carefully verifying the objects based on the experimental setup and our subjective knowledge. These annotated data samples were then used for training and evaluating the model, particularly focused on detecting the "Object" class. For the result analysis, we employed several widely accepted evaluation metrics to assess the model's segmentation performance; e.g.; Segmentation Dice Coefficient (DC), Intersection over Union (IoU), Pixel Accuracy (PA), Precision (PS), Recall (RS), F1 Score (F1S), Mean Absolute Error (MAE), Boundary IoU (BIoU), and BIoU/Boundary F1 Score (BS).

We performed two experiments to evaluate the U-Net model’s segmentation performance. In \textbf{Experiment 1}, we used the polar occupancy map with data preprocessing, including noise reduction techniques. In \textbf{Experiment 2}, we used the raw polar occupancy map directly from the sensor APIs without any types of preprocessing. Our hypothesis was that the preprocessed dataset would yield better performance due to the noise reduction, which would allow the model to focus more effectively on object features. As expected, the results showed that the preprocessed data significantly outperformed the raw data, achieving an overall accuracy of 93\%. In contrast, the model trained on raw data reached an accuracy of 88\%. This demonstrates the importance of preprocessing in improving segmentation accuracy.

\begin{table}[htbp]
    \centering
    \caption{Performance metrics for U-Net model segmentation on preprocessed and raw data}
    \label{tab:3}
    \begin{tabularx}{\textwidth}{c|X|c|c}
    \toprule
    \textbf{Metric} & \textbf{Definition} & \textbf{Experiment 1 (Preprocessed Data)} & \textbf{Experiment 2 (Raw Data)} \\
    \midrule
    \textbf{DC} & Measures the overlap between the predicted segmentation and the ground truth. & 0.92 & 0.86 \\
    \textbf{IoU} & Ratio of the intersection of predicted and ground truth regions to their union. & 0.89 & 0.82 \\
    \textbf{PA} & The ratio of correctly classified pixels to the total number of pixels. & 0.93 & 0.88 \\
    \textbf{PS} & The ratio of correctly predicted positive pixels to all predicted positives. & 0.91 & 0.85 \\
    \textbf{RS} & The ratio of correctly predicted positive pixels to all actual positives. & 0.94 & 0.87 \\
    \textbf{F1S} & The harmonic mean of precision and recall, offering a balanced evaluation of pixel-wise predictions. & 0.93 & 0.86 \\
    \textbf{MAE} & Represents the average of the absolute differences between predicted and actual pixel values. & 0.07 & 0.12 \\
    \textbf{BIoU} & Evaluates the ratio of intersection over union specifically on object boundaries. & 0.88 & 0.83 \\
    \textbf{BS} & A harmonic mean of precision and recall specifically at object boundaries, assessing boundary accuracy. & 0.90 & 0.84 \\
    \bottomrule
    \end{tabularx}
\end{table}

Table~\ref{tab:3} shows that Experiment 1, which involved preprocessing the data, outperformed Experiment 2 (raw data) in all metrics. For instance, Experiment 1 achieved a PA of 0.93 and a DC of 0.92, reflecting better overlap and more accurate pixel classification compared to Experiment 2, where PA was 0.88 and the DC was 0.86. The improvements in metrics such as IoU, F1S, and BIoU further demonstrate that the noise reduction and preprocessing allowed the model to focus more effectively on the true object boundaries, reducing false positives. The results confirm that preprocessing enhances the model's ability to handle sonar data and leads to higher segmentation performance, making it critical for achieving accurate object detection in complex environments. Moreover, we have used a basic architecture of U-Net, more complicated architecture will give better results, and these results can be further improved.

\section*{Response to Our Research Questions}

Our study aimed to explore key challenges in interpreting and processing sonar data for underwater object detection, particularly with low-cost devices like the Ping 360, while focusing on addressing gaps in existing research. Throughout our experiments, we encountered several challenges that directly relate to our RQs, and our findings provide valuable insights into the complexities of sonar data and its use in object detection. 

One of the primary challenges we investigated was interpreting navigation sonar data in cluttered or reflective environments, as outlined in RQ 1. The sonar data we collected often exhibited issues like acoustic shadows, surface reflections, and noise, especially in environments with multiple objects or reflective surfaces like ceramic tiles. For example, in Experiment 8, the placement of multiple objects generated overlapping acoustic shadows, complicating object detection. These shadows, caused by objects blocking the sonar waves, often concealed other objects behind them, making it difficult to differentiate between objects solely based on sonar imagery. Furthermore, surface reflections introduced additional noise, particularly in the areas beyond the pool boundaries, as demonstrated in the polar occupancy maps. These factors, which are common in real-world underwater environments, reveal that accurately interpreting sonar data requires a nuanced understanding of acoustic phenomena, something that is often overlooked in traditional object detection studies that rely heavily on image-based methods. 

RQ 2 focused on the effectiveness of manual annotation in improving object detection when used with advanced AI models, such as U-Net. Through our study, we found that while manual annotation can significantly enhance the performance of AI models in sonar data analysis, the nature of sonar data requires more than just treating it as a standard computer vision problem. Unlike optical images, sonar data reflects acoustic properties, where ambiguities such as acoustic shadows, noise, and surface reflections are prevalent, making it difficult to directly apply computer vision techniques designed for visual imagery. For example, in Experiment 8, the sonar data clearly identified obstacles like the PVC pipe and bucket, but without additional human expertise and context, accurately distinguishing between these objects remained challenging. Sonar images often lack the clarity and detail present in visual images, meaning that bounding boxes/segmentation masks drawn during annotation may only tell us where an object or obstacle is, but not precisely what that object is. Most existing literature treats sonar outputs like optical images and applies standard AI models for segmentation or detection. However, this approach oversimplifies the complexities of sonar data, ignoring factors like acoustic distortions and the environmental context in which the sonar is used. Therefore, while advanced AI models like U-Net can certainly detect objects, relying solely on these models without considering the unique characteristics of sonar data is insufficient. Successful object detection requires a more integrated approach, combining manual annotation with domain expertise and preprocessing techniques that account for the acoustic properties of sonar. In this context, our findings suggest that sonar data cannot be treated like traditional computer vision problems, and manual annotation alone—though effective—is not enough unless it is guided by a deep understanding of the underwater environment and the limitations of sonar imaging.

Lastly, RQ 3 asked whether low-cost sonar devices like the Ping 360, typically used for navigation, could be repurposed for more complex underwater object detection. Our findings suggest that while the Ping 360 sonar can be adapted for object detection, there are significant limitations to its use in complex environments. The device performed adequately in detecting simple objects at close ranges, such as PVC pipes and buckets, particularly when there were no overlapping objects to create acoustic shadows. However, as the experiments became more complex—such as when objects were aligned closely or when surface reflections interfered—the sonar's ability to distinguish between objects diminished. Additionally, the device’s limitations in terms of resolution and sensitivity to noise became apparent, particularly when attempting to detect smaller distant objects (less than 1 meter). While the sonar can be repurposed for basic object detection tasks, more sophisticated methods—such as advanced preprocessing, noise reduction, and expert-guided annotation—are necessary to achieve reliable results in real-world scenarios.

Therefore, our study highlights the complexities involved in interpreting sonar data, the importance of expert-guided manual annotation, and the potential but limited use of low-cost sonar devices like the Ping 360 for complex underwater object detection. By addressing these challenges and applying thoughtful preprocessing and expert knowledge, we were able to improve the performance of our object detection models, though the limitations of the sonar device itself remain a key factor in determining its effectiveness in more demanding environments. 

\section*{Conclusions}

This study provides the first detailed investigation into the feasibility of using the Ping 360 sonar device for complex underwater object detection. Our research was motivated by the affordability and open-source nature of the Ping 360 compared to higher-cost imaging sonar devices, making it an attractive option for budget-constrained projects. Through a series of controlled experiments, we evaluated the device’s behaviour in detecting objects under various conditions, focusing on factors such as surface reflections, object shadows, and effect of shallow water. We found that while the Ping 360 performs adequately in simple environments, its limitations become evident in cluttered or reflective underwater settings. Without thoughtful data pre-processing and manual annotation, the sonar’s raw output lacks the clarity necessary for accurate object detection, particularly in complex scenarios.

One of the key contributions of this study is the creation and release of a dataset consisting of raw sonar data collected using the Ping 360 sonar device \cite{DP1}. This dataset provides a valuable resource for future researchers who may not have access to the device, allowing them to explore its potential for object detection and develop novel AI-based approaches for sonar data interpretation. The dataset also underscores the opportunities of further data pre-processing and careful annotation, as these were critical in improving the performance of any AI based segmentation/classification model. Our findings highlight several challenges inherent in sonar data interpretation, including the effects of surface reflections and acoustic shadows, which often obscure objects and complicate detection. In scenarios where multiple objects were present, these factors made it difficult for the sonar to reliably differentiate between targets without human intervention. We demonstrated that the success of machine learning models, such as the U-Net segmentation algorithm, is highly dependent on the quality of the data and the accuracy of the annotations used for training. This emphasizes the need for expert knowledge and careful curation of training data. Overall, this work provides a foundation for future research to further explore and enhance the capabilities of affordable sonar technology in underwater applications.

\bibliography{main}

\section*{Declaration of generative AI and AI-assisted technologies in the writing process}

During the preparation of this work the author(s) used ChatGPT, and Quill Bot in order to refine the writing, in-terms of language. After using this tool/service, the author(s) reviewed and edited the content as needed and take(s) full responsibility for the content of the publication.

\section*{Data availability statement}

The dataset can be downloaded from the following link: \href{https://github.com/junayed/Sonar-Scan-Dataset-with-Ping-360}{https://github.com/junayed/Sonar-Scan-Dataset-with-Ping-360}.

\section*{Acknowledgement}

We would like to express our sincere gratitude to Bryan Bourdin and Ross Henderson for their valuable assistance during the data collection process. Their support were instrumental to the success of this work.

\section*{Author contributions statement}

M.J.H. and S.K. were responsible for the experimental design, hardware setup, and data interpretation. M.J.H. conceptualized the study, developed the methodology, conducted the formal analysis, and led the writing of the manuscript. M.J.H. also handled the data curation, visualization, and the development of the machine learning models. S.K. and A.R. provided valuable feedback by revising the initial draft of the manuscript. M.A.S. contributed by providing project support, and assisted with manuscript revision. 

% \section*{Additional information}

% To include, in this order: \textbf{Accession codes} (where applicable); \textbf{Competing interests} (mandatory statement). 

% The corresponding author is responsible for submitting a \href{http://www.nature.com/srep/policies/index.html#competing}{competing interests statement} on behalf of all authors of the paper. This statement must be included in the submitted article file.

% \begin{figure}[ht]
% \centering
% \includegraphics[width=\linewidth]{Figs/stream.jpg}
% \caption{Legend (350 words max). Example legend text.}
% \label{fig:stream}
% \end{figure}

% \begin{table}[ht]
% \centering
% \begin{tabular}{|l|l|l|}
% \hline
% Condition & n & p \\
% \hline
% A & 5 & 0.1 \\
% \hline
% B & 10 & 0.01 \\
% \hline
% \end{tabular}
% \caption{\label{tab:example}Legend (350 words max). Example legend text.}
% \end{table}

% Figures and tables can be referenced in LaTeX using the ref command, e.g. Figure \ref{fig:stream} and Table \ref{tab:example}.

\end{document}